\documentstyle[11pt]{article}
\begin{document}

\begin{center}
\large \bf
 First-principles studies of modulated Co/Cu superlattices with strongly \\
and weakly exchange-biased Co-monolayers leading to a ferrimagnetic ground
 state
\end{center}
\vspace{1cm}

{\large S.~Krompiewski$^\dagger$, F.~S\"uss* and
 U. Krey*{\footnote{corresponding author, FAX xx49 941 943 4544,
e-mail:
krey@rphs1.physik.uni-regensburg.de}}}
\vspace{1cm}

\normalsize {\it $^\dagger$ Institute of Molecular Physics,
 P.A.N., Smoluchowskiego 17, PL-60-179 Pozna\'n, Poland \\
${}^*$ Institut f\"ur Physik II, Universit\"at Regensburg,D-93040
 Regensburg, Germany}
 \vspace{1cm}

 Received ........... December 1995

PACS. 75.50R -- Magnetism in interface structures (incl.~layers and
superlattice structures). \newline
PACS. 75.30E -- Exchange and superexchange interactions.\newline
PACS. 75.70F -- Magnetic ordering in multilayers.

 \vspace{1cm} \hrule \vspace {0.5cm} {\bf Abstract.}

First-principles calculations have been performed in order to
determine effective exchange integrals between {\it strongly} and {\it
weakly} exchange-biased Co monolayers in the modulated $CoCu_2/CoCu_n$
superlattices. For $3\le n\le 6$ it has been found that the respective
exchange integrals have opposite signs and differ for $n\ne 4$ from
each other by one order of magnitude and for $n=4$ still by a factor
of $\sim 1.7$. The obtained phase diagram,
 with all the relevant magnetic
phases, shows that for the $n$-values considered the ground state
configuration is ferrimagnetic.  \vspace{0.5cm}\hrule \vspace {0.5cm}

Magnetic multilayers based on magnetic transition metals with
nonmagnetic spacers have been intensively studied for almost five
years now, after it was realized that they reveal unusual oscillatory
behaviour of the exchange coupling and magnetoresistance
\cite{l:par1}.  The oscillatory phenomena have a universal character,
do not depend drastically on the kind of metals involved \cite{l:par2}
and occur both with the spacer thickness as well as magnetic slab
thickness variations \cite{l:brun,l:blo1,l:kr1}.

 Recently a great deal
of attention has been attracted by exchange-biased systems
$AF/F_1/S/F_2$ \cite{l:dien}, with one ferromagnetic slab ($F_1$)
strongly coupled to an antiferromagnet ($AF$) (e.g. $MnFe$, $CoO$ or
$NiO$) and the other slab ($F_2$) -- almost free -- only weakly
coupled to the first one via the spacer ($S$).  Systems of this type
are not only interesting for fundamental aspects but may also be
applied in future magnetic recording devices. It is possible to
replace an exchange-biasing antiferromagnet by a trilayer, $AF
\longrightarrow ferromagnet1/spacer/ferromagnet2$, provided the
thickness of the spacer is chosen such as to ensure strong
antiferromagnetic coupling of the two ferromagnetic layers
\cite{l:blo2}.

In an attempt to get more insight into the nature of exchange coupling
and possible magnetic phases which may appear, we have studied
systematically by the spin-polarized {\it ab initio LMTO-ASA} method
(linearized muffin-tin orbitals, atomic sphere approximation,
scalar-relativistic version, see [5]) the series of multilayers with
supercell \newline $Cu_2Co^{(2)}Cu_2Co^{(1)}Cu_2Co^{(2)}Cu_nCo^{(3)}Cu_n$ of
the (001) face-centred tetragonal structure (i.e. $Co$ is grown
epitaxially on $Cu$).  In these sytems the monolayers $Co^{(1)}$
couple strongly antiferromagnetically with $Co^{(2)}$, while the
$Co^{(3)}$ monolayers are only weakly coupled (for $n>2$). In contrast
to other studies our systems are infinite multilayers, and the
sublayer $Cu_2Co^{(1)}Cu_2$, which acts as a strong antiferromagnetic
bias on the $Co^{(2)}$ monolayers, is periodically repeated.

We have built our structural models in a similar way as in our earlier
papers \cite{l:kr1,l:kr2}, in particular the in-plane atomic spacings
are assumed to be equal to those of the $fcc-(001)\, Cu$ with the
lattice constant $a=3.615$ \AA. The main task has been to determine
both strong as well as weak exchange couplings from total energy band
calculations for all the relevant spin configurations, namely:
\begin{eqnarray*}
\begin{array}{lll}
(i) & Co\downarrow Cu_2 Co\uparrow Cu_2 Co\downarrow Cu_n Co\downarrow
Cu_n & ([\downarrow\uparrow\downarrow,\downarrow]) \, , \\
(ii) & Co\downarrow Cu_2 Co\uparrow Cu_2 Co\downarrow Cu_n Co\uparrow
 Cu_n & ([\downarrow\uparrow\downarrow,\uparrow]) \, , \\
(iii) & Co\uparrow Cu_2 Co\uparrow Cu_2 Co\uparrow Cu_n Co\uparrow
Cu_n & ([\uparrow\uparrow\uparrow,\uparrow]) \, . \\
\end{array}
\end{eqnarray*}

Obviously, since in the present studies no anisotropy is included, all
the systems are spin-rotationally invariant, and there is no
distinction whatsoever between the above mentioned configurations and
the ones with all the spins rotated simultanously by an arbitrary
angle.  After having computed the total energies of the above
configurations ($E_1, E_2$ and $E_3$) the corresponding $Cu$-mediated
exchange coupling integrals have been directly found from
\begin{eqnarray}
j &=& \frac{1}{4} (E_2-E_1) /A \, , \\
J &=& -\frac{1}{4} (E_3-E_1) /A \, ,
\end{eqnarray}

where $A$ is the  cross-section area of the unit supercell and $E_i$
are the energies per supercell in the above-mentioned
states. Furthermore,
 one factor of $\frac{1}{2}$ in eqs. (1) and (2) comes from the fact
that there are two thick spacers (related to the small exchange
coupling $j$) and two thin spacers (related to the large one,
$J$), whereas the other factor of $\frac{1}{2}$ results from the spin flip
process according to the well known Heisenberg interaction energy per
''bond'' $<ij>$:

\begin{equation} E_{<ij>} = -J_{ij}
\frac{\vec{S}_i\cdot\vec{S}_j}{\mid \vec{S}_i\mid \mid \vec{S}_j \mid}
\, .  \end{equation}

The main result of the present letter is presented in Fig. 1, where
the calculated exchange integrals are visualized. It can be seen that
the computed couplings $j$ and $\mid J \mid$ oscillate with the $Cu_n$
spacer thickness in a similar way. The oscillations of the strong
coupling $J$ have a large negative bias (i.e.~they favour antiparallel
ordering; the absolute value is plotted!) and have a much higher
amplitude the oscillations of the weak coupling $j$, which remains
positive (i.e.~ferromagnetic) for $3\le n \le 6$.  The coupling $j$
has the correct order of magnitude (except of the case $n=4$) when
compared with experimental results on similar systems
(e.g. \cite{l:blo2}).

This means that the exchange-biasing slab $Cu_2Co^{(1)}Cu_2$
influences the coupling $j$ (between $Co^{(2)}$ and $Co^{(3)}$ via
$Cu_n$) and reduces it in a substantial way, since in our previous
studies with no biasing slab the exchange coupling in $CoCu_n$
multilayers was roughly 10 times larger than the present $j$-value
(see \cite{l:kr2,l:kr1} and comments therein).

The $j$- and $\mid J \mid$-curves in Fig. 1 separate various magnetic
phases.  As $j$ never crosses zero it means that at least for $3\le
n\le6$ and vanishing external magnetic field, the ground state of the
multilayers under consideration is the ferrimagnetic state
$[\downarrow\uparrow\downarrow ,\downarrow]$ (up to spin-rotational
equivalence).  One may predict that in this state the resistance of
the multilayers, in comparison with that in the saturated state
$[\uparrow\uparrow\uparrow,\uparrow]$, will be considerably increased
for the {\it CPP} (current perpendicular to the plane) geometry, and
moderately increased for the {\it CIP} (current in plane) geometry
\cite{l:sche}. This magnetoresistance effect would be more spectacular
if the ground state configuration had totally compensated spins
$[\downarrow\uparrow\downarrow ,\uparrow]$ or if such a totally
compensated state could have been reached by applying a magnetic
field: The latter possibility could happen, if the spins of the
biasing $\downarrow\uparrow\downarrow$-unit were strongly pinned by a
substrate or some strong internal anisotropy such that by application
of an external field in $\uparrow$-direction the transition from the
ferrimagnetic $[\downarrow\uparrow\downarrow,\downarrow]$ state to the
{\it totally spin-compensated} state
$[\downarrow\uparrow\downarrow,\uparrow]$ would happen earlier than
the transition to the {\it totally reversed} state
$[\uparrow\downarrow\uparrow,\uparrow]$.  However it would go beyond
the scope of the present work to discuss all possible scenarios in
detail.

\vglue 0.2 truecm
In conclusion, our ab-initio calculation with a spin-polarized
LMTO-ASA method for the possible magnetic
configurations of modulated $CoCu_2/CoCu_n$ superlattices has shown the
simultaneous presence of {\it strongly} and {\it weakly}
exchange-biased $Co$ monolayers for $3 \le n \le
6$.
  It has been found that the strong coupling across two $Cu$
monolayers is antiferromagnetic and much larger (namely by one order
of magnitude for $n\ne 4$, and still by a factor of $\sim 1.7$ in the
case of $n=4$) than the coupling across the thicker spacer $Cu_n$ with
$3\le n\le 6$, which is ferromagnetic. The ground state of the systems
considered is then the ferrimagnetic state (i) from above, corresponding to
the lowest configuration sketched in the figure. Implications of these
findings for the magnetoresistance behaviour, i.e.~for possible
applications, have also been briefly mentioned.  \vspace{0.5cm}

{\bf Acknowledgements}
\vspace{0.5cm}

We would like to thank Profs. F. Stobiecki and G. Bayreuther for valuable
discussions.
This work has been carried out under the grant no. 2 P 302 005 07 (SK), and
the bilateral project DFG/PAN 436 POL (UK and SK). We also thank the  Pozna\'n,
Munich and Regensburg Computer Centres for computing time.

\newpage

{\Large \bf Figure Captions}
\vspace{1cm}

Fig.1: Phase diagram for the modulated $CoCu_2CoCu_2CoCu_nCoCu_n$
superlattices with strongly and weakly exchange-biased Co
monolayers: The strong antiferromagnetic exchange coupling ($J$;
dotted line) acts between two Co monolayers separated by just two $Cu$
monolayers ($Cu_2$), whereas the weak ferromagnetic coupling ($j$;
dashed line) occurs across $Cu_n$ with $3\le n\le 6$. According to
eqs.~(1) and (2), $j$ and $|J|$ are $\propto (E_2-E_1)$ and
$(E_3-E_1)$, respectively, where $E_3$, $E_2$ and $E_1$ refer to the
spin configurations sketched in the figure from above.
%
%
\newpage
\input epsf
\begin{figure}[htb]
\epsfxsize=16cm
\epsfbox{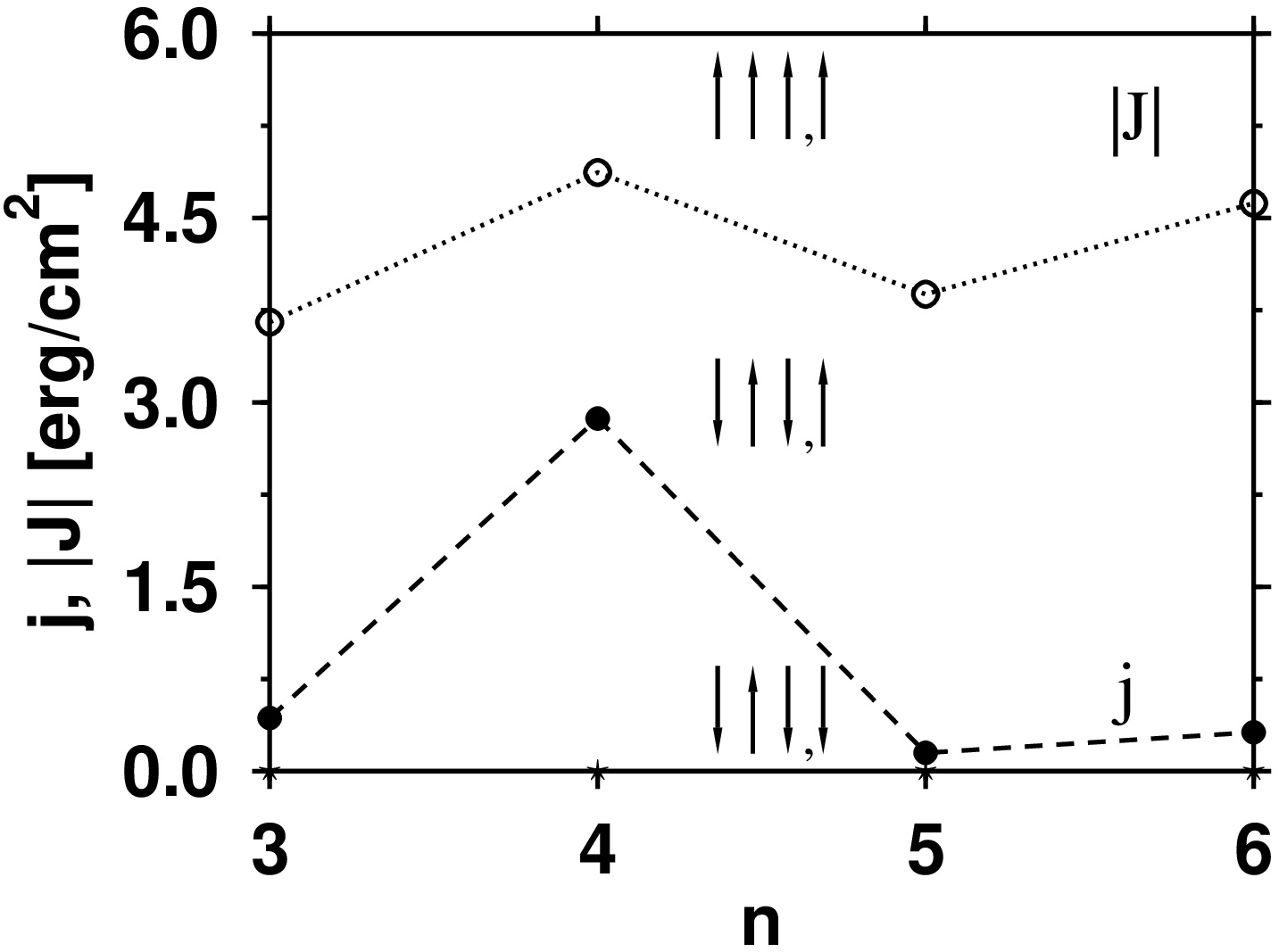}
\end{figure}
\end{document}